\documentclass[aps,prl,twocolumn,groupedaddress,amsfonts,amssymb]{revtex4-1}
\usepackage{multirow}
\usepackage{amssymb,amsmath}
\usepackage[dvips]{graphicx}
\usepackage{epsfig}

\newcommand{\dm}{$\mathbf{d}$}

\begin{document}

\title{Factors influencing the distribution of charge in polar nanocrystals}

\author{Philip W. \surname{Avraam}}
\author{Nicholas D. M. \surname{Hine}}
\author{Paul \surname{Tangney}$^{*}$}
\author{Peter D. \surname{Haynes}}

\affiliation{Department of Physics and Department of Materials,
  Imperial College London, Exhibition Road, London SW7~2AZ, United
  Kingdom}

\date{\today}

\begin{abstract}
We perform first-principles calculations of wurtzite GaAs nanorods
  to explore the factors determining charge distributions in polar
  nanostructures. We show that both the direction and magnitude of the dipole 
  moment~\dm{} of a nanorod, and its electic field,
  depend sensitively on how its surfaces are 
  terminated and do not depend strongly on the spontaneous polarization of the underlying lattice. 
  We identify two physical mechanisms by which~\dm{} is controlled by the surface termination,
  and we show that the excess charge on the nanorod ends is not
  strongly localized.
  We discuss the implications 
  of these results for tuning nanocrystal properties, and for their growth and assembly.
  
\end{abstract}


\maketitle


Nanocrystals and nanorods are valued for their unique electronic and
optical properties which differ substantially from
bulk materials of the same composition~\cite{smallisdifferent}. 
They are being exploited in a host of applications 
(including imaging in biology~\cite{X.Michalet01282005},
light-emitting diodes~\cite{NirTessler02222002}, lasing
devices~\cite{kazesetal} and solar cells~\cite{Wendy})
that continues to grow in number and
diversity as we gain finer control over their properties~\cite{Nieetal}. This
requires greater understanding of how those properties depend upon
size, shape, internal structure and chemical environment.
The distribution of charge within and on a nanorod
plays an important role in determining its physical properties:
nanostructures with large dipole moments (\dm) are sources of large
electric fields which, internally, affect their optical properties
and, externally, affect their interactions with their surroundings,
thereby influencing both their growth and assembly into
superstructures~\cite{shevchenko}.
The charge distribution is related to the chemical environment and to the 
conditions of synthesis, however this relationship is not well understood. 
In this Communication we present first-principles
calculations that provide substantial insight into how charge is
distributed within a nanocrystal and we identify the most important
factors determining this distribution.
Our findings can be used to inform the choice of synthesis conditions
appropriate for the design of nanorods with specific physical
properties.

Much theoretical and experimental work on nanorods has focussed on the
magnitude and origin of their dipole moments, with somewhat contradictory
results. Some studies at\-trib\-ute large dipole moments to the
non-centro\-symmetric nature of the wurtzite
structure~\cite{noncentro} which is associated with an intrinsic
spontaneous polarization -- an interpretation that is widely
accepted~\cite{talapin}. However a theoretical study has revealed a strong enhancement 
of the polarity of nanorods compared with thin films of the
same length along the $[0001]$ direction ~\cite{tsai} and experimental 
observations by Shim and Guyot-Sionnest~\cite{shim:6955} show that ZnSe nanocrystals with the cubic zincblende structure can exhibit moments of similar magnitude 
to wurtzite CdSe. Both studies cast doubt on the relevance of crystal symmetry.
Other studies have attributed
importance to different factors, including nanocrystal
shape~\cite{Shanbhag:2006ix}, molecular passivation of
surfaces~\cite{Shanbhag:2006ix, rabani:1493}, surface
reconstruction~\cite{rabani:1493, cicero} and a piezoelectric effect caused by
strain at the nanocrystal surfaces~\cite{huong:1769}. Finally, an
electrostatic force microscopy study of
CdSe nanorods observed no
dipole moment in the samples studied~\cite{PhysRevLett.92.216803}.

Computational modelling of nanorods gives us the control necessary to
disaggregate the factors contributing to the dipole moment in a way not 
possible experimentally. Until recently, nanostructures of realistic sizes 
have been beyond the reach of accurate quantum-mechanical
methods. However, developments in linear-scaling
density-functional theory (DFT) methods have now made possible the
simulation of nanostructures comprising thousands of atoms with high
accuracy. 
We take advantage of these methods, as implemented in our ONETEP 
code~\cite{skylaris:084119,PhysRevB.83.195102}, to accurately simulate polar semiconductor nanorods of realistic sizes.
For our investigation, we choose nanorods of wurtzite GaAs
because GaAs has a relatively low computational cost while having the
essential features that
bestow all polar nanorods (e.g. CdSe, ZnO) with an asymmetric distribution of charge, namely, there is a degree
of ionicity to the bonding and the crystal structure lacks inversion symmetry.
This allows us to access the extensive size regime in 
which a nanorod's dipole moment increases linearly with its length and width.

Our results highlight the importance of surface chemistry to the
distribution of charge in polar semiconductor nanorods and show that
the symmetry of the corresponding bulk crystal structure can play a
much less important role than has often been assumed. Indeed, for some
surface terminations, \dm{} can be in the opposite direction to that
suggested by the spontaneous polarization of the bulk crystal. We show
that excess charge on the ends of a nanorod can be highly delocalized,
meaning that internal electric fields are non-uniform and that some simple
models of electrostatic interactions between nanoparticles may be
overly simplistic. We explain the relationships that we find between
the surface terminations of a nanorod and its dipole moment in terms of the
electronic structure. Finally, we show that our findings are robust
when the atomic structure is allowed to relax.

\begin{figure}[ht]
\includegraphics[width=86mm]{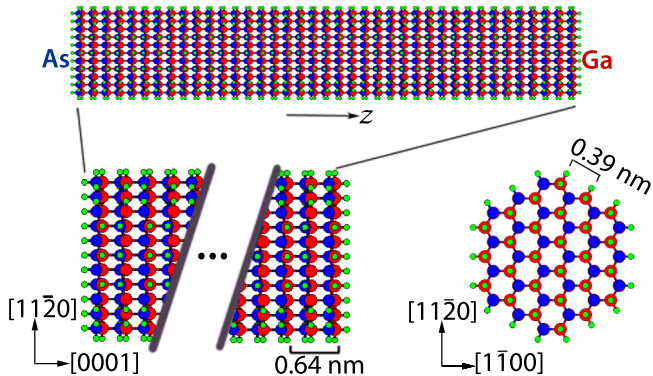}
\caption{(Color online)An unrelaxed GaAs nanorod with Ga, As, and H atoms colored red (medium grey), blue (dark grey), and green (light grey), respectively. 
The Ga and As terminated ends are indicated.}
\label{fig:rods}
\end{figure}

The ONETEP code uses DFT in a formulation equivalent to the plane-wave
pseudopotential method~\cite{RevModPhys.64.1045}. 
Initially we model unrelaxed, stoichiometric nanorods of wurtzite (w-)
GaAs. 
Our nanorods are `grown' parallel to the wurtzite $c$-axis with a
length $\sim 12$~nm and hexagonal cross-sections of width $\sim 2$~nm:
an example is shown in Fig.~\ref{fig:rods}.
%
%
Six different nanorods consisting of 2106--2862 atoms were created to
represent a variety of lateral ($\parallel c$) and polar ($\perp c$) 
surface terminations, either bare or saturated by chemisorption.
They are labelled H/H, H/B, H/P, B/H, B/B, B/P, to indicate how the
lateral/polar surfaces are terminated, respectively. H signifies
termination by hydrogen atoms and B signifies a bare surface. P
denotes termination by
pseudo-atoms~\cite{PhysRevB.71.165328}, each with one electron and
fractional nuclear charges of $+\textstyle{\frac{5}{4}}$ and
$+\textstyle{\frac{3}{4}}$ for passivating surface Ga and As atoms
respectively. By approximating the electronegativities of As and Ga
they form bonds with the surface atoms of similar character to those
found in the bulk and render the III-V semiconductor surfaces
electronically inert~\cite{PhysRevB.71.165328} without contributing
any net charge to them. A H atom, on the other hand, passivates a
surface dangling bond but also contributes a net negative (positive)
charge of magnitude $\textstyle{\frac{1}{4}}$ to the surface when it
binds to a Ga (As) atom. In all of these models symmetry dictates that only
the longitudinal component of \dm{}, $d_z$, is non-zero. 

A plane-wave energy cut-off of 450~eV, with a local orbital
radius~\cite{0953-8984-17-37-012} of 0.53~nm for each atomic species
is found to be sufficient to converge all properties of interest. The
density kernel is not truncated so that metallic and insulating
structures are treated on an equal footing.
 We use norm-conserving pseudo-potentials with non-linear core
corrections, and 3d electrons frozen into the core.
Exchange and correlation are treated within the local density
approximation.
To eliminate interaction between a nanorod and its periodic images we
have used a cylindrically truncated Coulomb
interaction~\cite{PhysRevB.73.205119}.

Table~\ref{tab:Dipole-moments} shows that both the magnitude and
direction of \dm{} depend critically on the chemistry of both
the lateral and polar surfaces and therefore cannot be dominated by the spontaneous
polarisation of the wurtzite lattice.
We have calculated this quantity for bulk w-GaAs to be 0.005~C/m$^2$, implying a
contribution to $d_z$ for nanorods of our size of +62~D, if the
polarizations are similar~\cite{cicero}. H/B and B/B both show positive $d_z$,
meaning that the Ga-terminated end carries a net positive charge,
while B/P has a very small $d_z$ and the other rods exhibit negative $d_z$. These observations
suggest that the synthesis conditions of nanorods and their chemical
environments must play crucial roles in determining \dm{}, insofar as
they affect the coverage of the surfaces with adsorbates.

\begin{table}[t]
\caption{\label{tab:Dipole-moments} Dipole moment $d_z$, net charge of
  the left-hand half $Q_{\text{L}}$ and electric field at the
  mid-point $E_{\text{m}}$ of the nanorods.}
\begin{tabular}{lr@{.}lr@{.}lr@{.}lr@{.}lr@{.}lr@{.}l} \hline \hline
Nanorod & \multicolumn{2}{c}{H/H} & \multicolumn{2}{c}{H/B} &\multicolumn{2}{c}{H/P} &\multicolumn{2}{c}{B/H} &\multicolumn{2}{c}{B/B} &\multicolumn{2}{c}{B/P} \\ \hline
$d_z$ (D) & \multicolumn{2}{c}{-614} & \multicolumn{2}{c}{+330} & \multicolumn{2}{c}{-531} & \multicolumn{2}{c}{-235} & \multicolumn{2}{c}{+125} & \multicolumn{2}{c}{+41} \\
$Q_{\text{L}}$ (\textit{e}) & +1&00 & -0&56 & +0&95 & +0&39 & -0&18 & -0&08 \\
$E_{\text{m}}$ (V/nm) & -0&100 & +0&050 & -0&105 & -0&030 & +0&013 &
+0&005 \\ \hline \hline
\end{tabular}
\end{table}

\begin{figure}
\includegraphics[width=86mm]{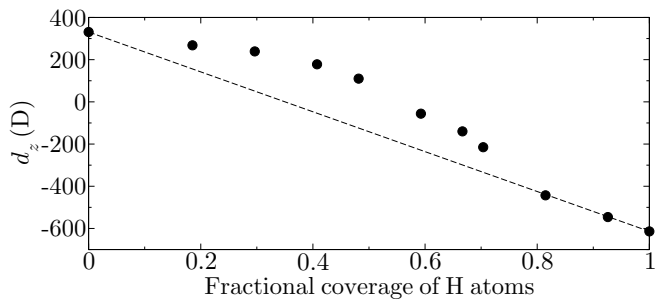}
\caption{Dipole moment $d_z$ as a function of H atom coverage of the polar surfaces for a nanorod with fully H-terminated lateral surfaces.}
\label{missing}
\end{figure}

Figure~\ref{missing} shows the calculated $d_z$ of nanorods with fully
H-terminated lateral surfaces, but with a varying coverage of H atoms
on the polar surfaces. Each point represents a single sample chosen at
random from the ensemble of nanorods with a given coverage. 
It is clear that $d_z$ tends to decrease significantly as the hydrogen
coverage increases and that \dm{} changes direction around 56\%
coverage. This means that if one could control the degree of coverage
of the polar surface (e.g. by varying the temperature, pressure, or the chemical potentials of
the various species during synthesis)
one could vary $d_z$ over a wide range of
values. At $T=0$~K, we have calculated that over the range of realistic hydrogen
chemical potentials $\mu_{\text H}$ for which water molecules are stable, the
thermodynamically stable hydrogen coverage, namely that which minimises
$E-\mu_{\text H} n_{\text H}$, goes from 0 to around 70\%. In principle, this therefore
allows access to values of $d_z$ of between -200~D and +330~D.
Even at full coverage, if there were competing adsorbing
species, we suggest that $d_z$ could be tuned by varying their
proportions. Indeed, terminating each of the polar surfaces with 13 H
and 14 pseudo-atoms yields $d_z=-552$~D, 
between H/H and H/P.

\begin{figure}[ht]
\includegraphics[width=86mm]{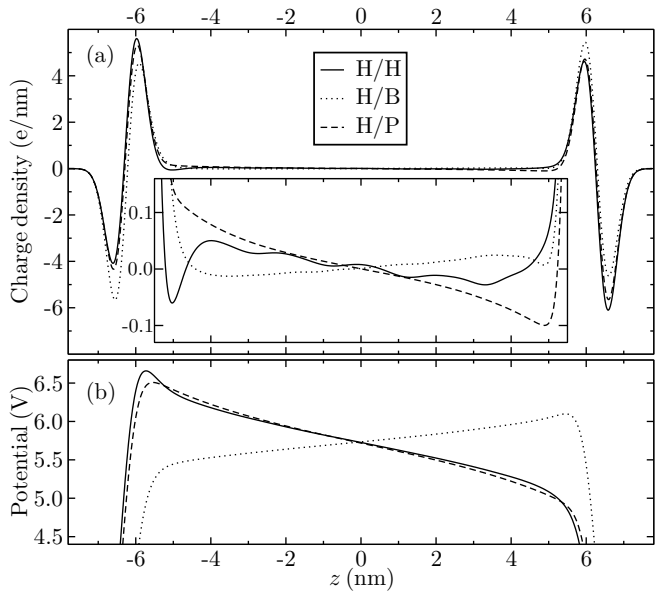}
\caption{\label{fig:denpot}(a) Charge density and (b) electrostatic
  potential as a function of position $z$ along the long axes of the
  H/H, H/B, and H/P nanorods. Both
  quantities are integrated over the plane perpendicular to the long
  axis and smoothed on the length scale of a wurtzite unit cell.}
\end{figure}

In Fig.~\ref{fig:denpot}(a) we plot the function
\[
\tilde{\rho}(z)=\frac{1}{L\sqrt{2\pi}}\int\int\int\rho(x^{\prime},y^{\prime},z^{\prime})\,\mathrm{e}^{-\frac{(z-z^{\prime})^{2}}{2L^{2}}}\,\mathrm{d}x^{\prime}
\, \mathrm{d}y^{\prime} \, \mathrm{d}z^{\prime} \:,
\]
which is the laterally-averaged charge density profile along the
length of the nanorod (the $z$ direction in this work, $\parallel c$),
convolved with a Gaussian of standard deviation 
$L=0.32$~nm
in the $z$ direction.
This smooths out the large variations in the density on the length
scale of a unit cell,
revealing
how excess
charge is distributed along the length of the nanorod: clearly it is
spread over several nm from the ends of the nanorods. The amount of
excess charge on the left-hand (As-terminated) end, $Q_{\text{L}}$, is shown
in Table~\ref{tab:Dipole-moments}. This has been
calculated by integrating the quantity $\tilde{\rho}(z)$ up to the
middle of the nanorod.

The observed spread of the excess charge
could be important for models of nanoparticle self-assembly that
usually treat nanocrystals as point dipoles or assume that
excess charge is perfectly localized on the polar surfaces~\cite{shevchenko,talapin}
These assumptions would lead to quantitatively, or even
qualitatively, incorrect results at short distances. Furthermore, the
delocalization of charge suggests that nanorods may be highly
polarizable, which could significantly affect the interactions between
nanorods.

Figure~\ref{fig:denpot}(b) shows the smoothed potential as a function
of position along three different nanorods and
Table~\ref{tab:Dipole-moments} gives the values of the electric field
at the centre of each nanorod, $E_{\text{m}}$.  It is clear
that the internal electric fields are not
uniform but stronger near the polar surfaces than in the middle. The
voltage drops we have observed are of order 0.1~V/nm, reaching as high
as 0.6~V/nm near the polar surfaces of H/H. These fields are of
similar magnitude to those observed in strained quantum
wells and are expected to significantly affect
optical absorption frequencies, selection rules and carrier
recombination rates.

\begin{figure*}
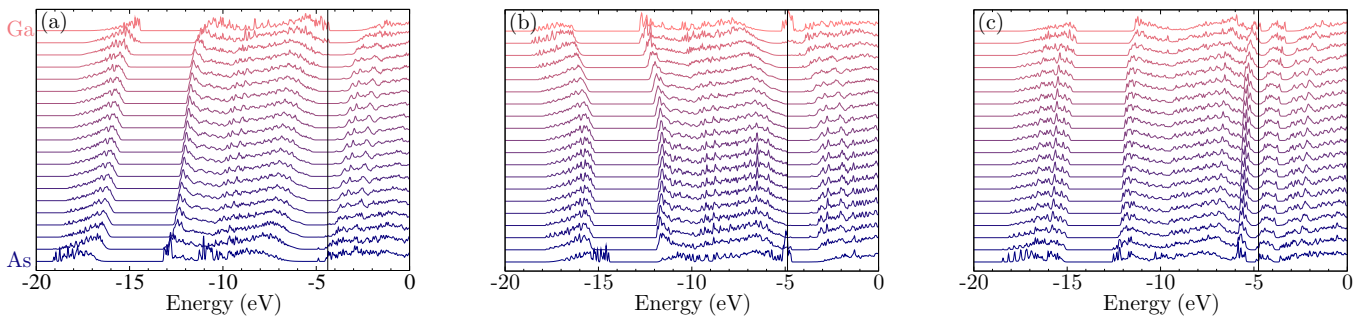

\includegraphics[scale=0.23]{figure4a.eps} 
\hfill
\includegraphics[scale=0.23]{figure4b.eps}
\hfill
\includegraphics[scale=0.23]{figure4c.eps} 
\caption{\label{fig:LDOS}(Color online) Local densities of states for three nanorods:
  (a) H/H; (b) H/B; (c) B/H. The global Fermi level is plotted in each
  case. Red (light grey) curves correspond to slabs close to the polar
  surfaces terminated by Ga and the blue (dark grey) curves correspond to slabs close to the polar
  surfaces terminated by As.}
\end{figure*}

Figure~\ref{fig:LDOS} shows the calculated `slab-wise' local densities
of states (LDOS) for nanorods H/H, H/B, and B/H.  We define a slab
LDOS as follows: each nanorod is nominally divided into 20 slabs
in the $z$-direction,
each consisting of four planes of atoms: two each of Ga and As. The
slab LDOS is the sum of the contributions to the total DOS from the
local orbitals centred on those atoms.
Superposing these slab LDOS, as in Fig.~\ref{fig:LDOS}, shows that the
electric field shifts the spectrum from slab to slab.  This shift has
the effect of smearing out the total DOS so that none of the nanorods
studied has an electronic energy gap despite individual slabs having
well-defined gaps.

Closer examination of Table~\ref{tab:Dipole-moments} reveals
two distinct trends: first, that all other things being equal,
terminated polar surfaces result in the most negative $d_z$, and bare
polar surfaces result in the most positive. Second, terminated lateral
surfaces result in \dm{} of larger magnitude than bare lateral
surfaces.

Figure~\ref{fig:LDOS} helps us to understand these trends: polar
surface dangling bond states on H/B can clearly be identified in
Fig.~\ref{fig:LDOS}(b) on the data sets associated with the first and
last slabs of the nanorod. The Ga (As) dangling bond states lie mostly
above (below) the Fermi level $E_{\text F}$, resulting in an excess of
electrons on the As-terminated end and a more positive $d_z$. In
contrast, Fig.~\ref{fig:LDOS}(a) reveals that both the Ga-H and As-H
bonding states on the polar surfaces of H/H lie below $E_{\text F}$,
resulting in an excess of electrons on the Ga-H terminated end and a
depletion on the As-H end. This is because the
$\textstyle{\frac{3}{4}}$ of an electron nominally available from the
Ga atom, and the one electron available from H are insufficient to
fully populate the Ga-H bond. Conversely, the
$\textstyle{\frac{5}{4}}$ electrons from As provide a surplus for the
As-H bond. Hence the formation of these bonds redistributes charge
between the two polar surfaces to produce a negative $d_z$. In real
systems, a mechanism is required to redistribute charge between two
distant end surfaces, such as the presence of a solvent capable of
mediating the transfer.

Regarding the second trend: comparison of Figs.~\ref{fig:LDOS}(a) and 
\ref{fig:LDOS}(c) shows that the nanorod with bare lateral surfaces (B/H) does not
exhibit a large local energy gap as H/H does due to the
presence of lateral surface dangling bond states.
A large \dm{} is associated with a large internal electric field and
slab-by-slab energy shift. However, for nanorods with lateral
surface states close to $E_{\text F}$, the electric field pushes
these states above $E_{\text F}$ at one end and below at the other
thereby reversing the charge build up on the ends and reducing the
field.
We conclude from these observations that a large local energy gap 
clear of surface states is a necessary condition for a large \dm{},
and the quenching of surface states on unstable lateral surfaces, by the
introduction of adsorbates, can be an important polarity 
enhancement mechanism in nanorods.

So far we have considered pristine nanorods. However, structural
relaxations
might be expected to play a large role for two reasons: some of the
surfaces studied above may be unstable and, given our arguments
above, reconstructions are likely to have a large effect; strain
caused by the surface may also induce significant charge
separation. Regarding the first concern, it has not been our intention
here to determine the stability of a particular surface, but rather to
investigate the link between termination and \dm{}, regardless of
stability. Secondly, to assess the effects of surface-induced strain
we performed structural relaxations on four nanorods of length $\sim
3.5$~nm, width $\sim 1.2$~nm and with different surface terminations
(b/b, h/b, h/h, b/h, following the same convention as
before). Table~\ref{tab:change in dipole moment} shows $d_z$
before and after structural relaxation: there is no qualitative change
in any of the values except b/h.  No dipole moments changed direction, and the ordering of
the magnitudes stayed the same. The increase in
$|\mathbf{d}|$ on b/h is attributable to the shifting of the
lateral surface dangling bond states away from $E_{\text F}$ on relaxation,
opening up the energy gap and, consistent with the argument above,
supporting a larger \dm{}.

\begin{table}
\caption{\label{tab:change in dipole moment}Configurations and dipole
  moments $d_z$ before and after structural relaxation of the four nanorods.}
\begin{tabular}{ccccr@{.}l r@{.}l r@{.}l r@{.}l} \hline \hline
\multicolumn{4}{l}{Nanorod} & \multicolumn{2}{c}{h/h} & \multicolumn{2}{c}{h/b} & \multicolumn{2}{c}{b/h} & \multicolumn{2}{c}{b/b} \\ \hline
\multirow{2}{*}{$d_z \Bigr\{$} & before & \multirow{2}{*}{$\Bigr\}$} & relaxation & -131&64 & +55&20 & -22&81 & +15&38 \\
& after & & (D) & -103&03 & +67&99 & -89&23 & +21&52 \\ \hline \hline
\end{tabular}

\end{table}

In summary, we find that for polar nanorods both the orientation and
magnitude of the dipole moment depend sensitively on the chemical 
terminations of both the polar end surfaces and the non-polar lateral 
surfaces. This sensitivity can overwhelm any contribution that may arise 
from the non-centrosymmetric crystal structure. The sensitivity to 
adsorbates arises in two main ways: adsorbates may be charged and 
therefore contribute directly to \dm{}; they also determine the stability
of surface electronic states which, in turn, determines the magnitude
of the internal electric field that a rod can sustain. When the
electric field
becomes large enough to shift the energy of an unoccupied state on one
end above an occupied state on the other,
a transfer of electrons between the ends occurs that lowers \dm{}, as
long as such a transfer can be facilitated by the environment. The
synthesis conditions and environment of a nanorod therefore play
crucial roles in determining both the ground state charge distribution
of a nanorod and whether or not it can reach this ground state.  We
also find that surface charge is not localized at the ends of the
nanorod but delocalized over several nanometers, also implying an non-uniform internal field. 
This has implications when considering the energetics of self assembly of polar
nanostructures.

This work was supported by EPSRC (UK) under Grant No.\ EP/G05567X/1,
the EC under Contract No.\ MIRG-CT-2007-208858 and a Royal Society
University Research Fellowship (PDH). All calculations were run on the
Imperial College HPC Service.\\
$^*$ Corresponding author; email: \url{p.tangney@imperial.ac.uk}.


\end{document}